\documentclass[reprint, amsmath,amssymb, aps,]{revtex4-2} 	

\usepackage[dvips]{graphicx}
\usepackage{dcolumn}								
\usepackage{bm}
\usepackage{bm,amsmath}
\usepackage{braket}
\usepackage{ulem}
\usepackage{here}

\usepackage{color}

\DeclareRobustCommand{\YMdel}{\bgroup\markoverwith{\textcolor[rgb]{0.1, 0.5, 0.1}{\rule[.5ex]{2pt}{0.4pt}}}\ULon}

\DeclareRobustCommand{\HHdel}{\bgroup\markoverwith{\textcolor[rgb]{0.1, 0.1, 0.9}{\rule[.5ex]{2pt}{0.4pt}}}\ULon}

\begin{document}

\preprint{APS/123-QED}

\title{A general method to construct mean field counter diabatic driving for a ground state search}

\author{Hiroshi Hayasaka$^1$} 
\email{hayasaka.hiroshi@aist.go.jp}
\author{Takashi Imoto$^1$} 
\email{takashi.imoto@aist.go.jp}
\author{Yuichiro Matsuzaki$^{1,2}$}
\email{matsuzaki.yuichiro@aist.go.jp,\\ present address: ymatsuzaki872@g.chuo-u.ac.jp}
\author{Shiro Kawabata$^{1,2}$}
\email{s-kawabata@aist.go.jp}
\affiliation{%
 $^1$Research Center for Emerging Computing Technologies (RCECT), \\
 National Institute of Advanced Industrial Science and Technology (AIST),\\
 1-1-1, Umezono, Tsukuba, Ibaraki 305-8568, Japan\\
 $^2$NEC-AIST Quantum Technology Cooperative Research Laboratory,\\
 National Institute of Advanced Industrial Science and Technology (AIST), Tsukuba, Ibaraki 305-8568, Japan
}%
\date{\today}
\begin{abstract}
The counter diabatic (CD) driving has attracted much attention for suppressing non-adiabatic transition in quantum annealing (QA). 
However, 
it can be intractable
to construct the CD driving in the actual experimental setup
 due to the
 non-locality of the CD dariving Hamiltonian and necessity of exact diagonalization of the QA Hamiltonian in advance. 
In this paper, using the mean field (MF) theory, we propose a general method to construct an approximated CD driving term 
consisting
of local operators.
We can efficiently construct the MF approximated CD (MFCD) term by solving the MF dynamics of magnetization using a classical computer.
As an example, we numerically perform QA with MFCD driving for the spin glass model with transverse magnetic fields.
We numerically show that the MF dynamics with MFCD driving is equivalent to the solution of the self-consistent equation in MF theory. Also,
we clarify that a ground state of the spin glass model with transverse magnetic field can be obtained with high fidelity compared to the conventional QA without the CD driving.
Moreover, 
we experimentally demonstrate our method by using a
D-wave quantum annealer and obtain the experimental result supporting our numerical simulation.
\end{abstract}
\pacs{Valid PACS appear here}
\maketitle

\section{\label{sec:level1}Introduction}
Quantum annealing (QA) is a computational method to obtain a nontrivial ground state by the adiabatic time evolution starting from a trivial ground state of the well-known Hamiltonian
\cite{Apolloni_1989, Finnila_1994, Kadowaki_1998, Farhi_2001, Farhi_2000, Arnab_2008}. 
The adiabatic theorem guarantees that,
if the annealing time scales as inversely
proportional to the square of the energy gap, the dynamics becomes adiabatic.

QA is of not only academic but also practical interest
such as quantum chemistry \cite{Xia_2017, Streif_2019, Genin_2019}, quantum state preparation \cite{Aspuru_2005, Veis_2014, Du_2010, Sugisaki_2022, Imoto2_2022}, combinatorial optimization problems, and database search \cite{Roland_2002}.   
In the last decade, QA was demonstrated with thousands of qubits in a programmable device
\cite{johnson2011quantum}. 
It has been reported that spin glass, Berezinskii-Kosterlitz-Thouless phase transition and $\mathbb{Z}_2$ spin liquid phase have been observed in the D-wave annealer \cite{Harris_2018, King_2018, Zhou_2021}. 
The quantum annealer has the potential to explore an exotic quantum phase where the classical computer can not be accessible.

If the QA Hamiltonian undergoes the
first order
quantum phase transition, 
it is impractical to implement QA because an exponentially long annealing time is required 
to be adiabatic \cite{Kato_1950, messiah_2014, Jansen_2007, Morita_2008, Amin_2009, Kimura_2022}. 
To circumvent this difficulty, various methods have been proposed \cite{Seki_2012, Seki_2015, susa2022nonstoquastic, Susa_2018, Grass_2019, Watabe2020, Karanikolas2020, Imoto_2021, Kadowaki_2023}. 
\begin{figure*}[t!]
\includegraphics[scale=0.30, bb=850 0 700 900]{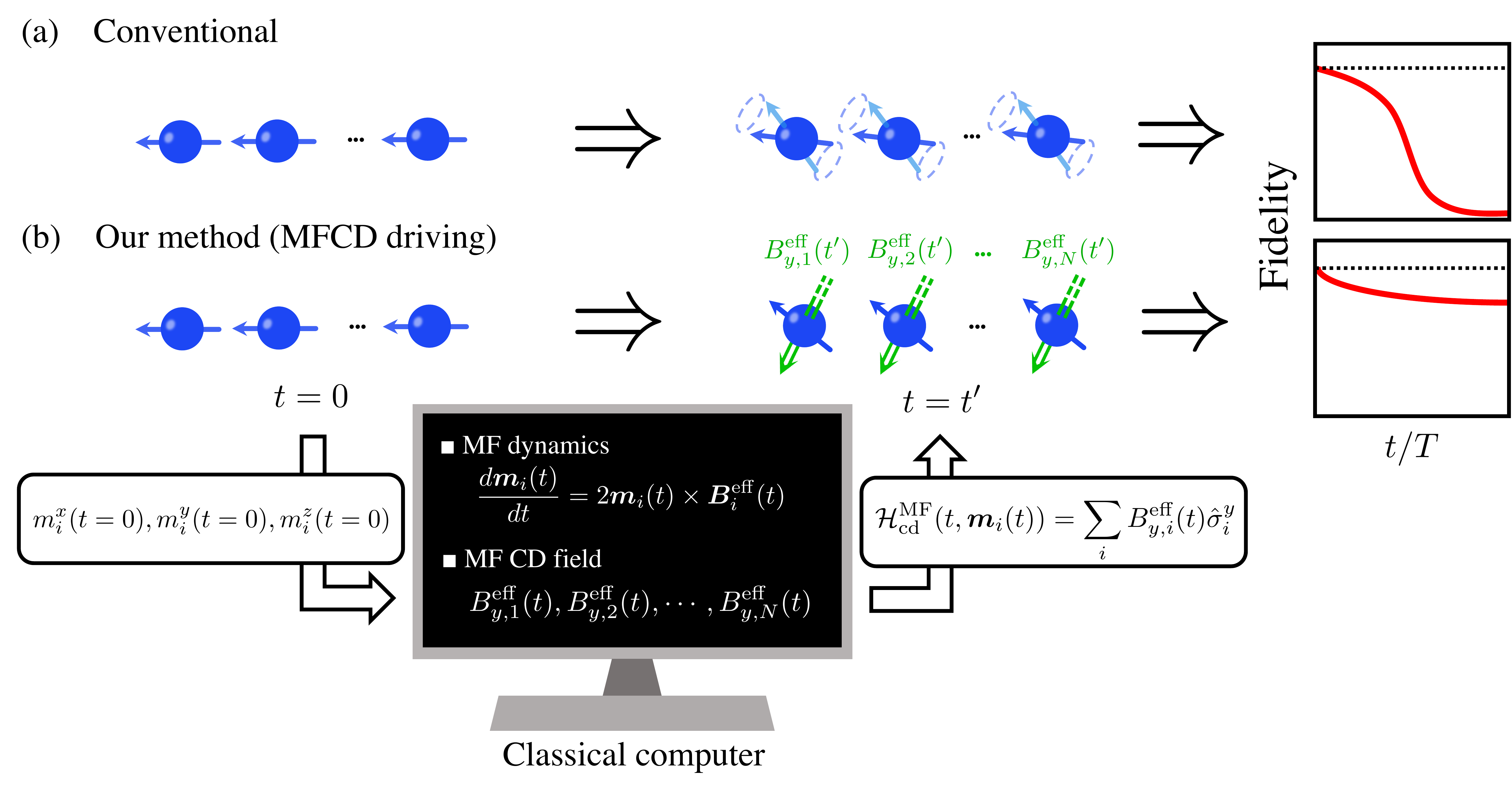}
\caption{A schematic of our method. (a) and (b) are the conventional QA and our method, respectively. 
(a) In conventional QA, the dynamics of magnetization does not track the instantaneous stationary state if annealing time is short, which causes precession of magnetization. 
(b) In our method, we calculate the dynamics of magnetization subjected to the effective magnetic field of MF theory in a classical computer in advance. Using this magnetization, we construct MFCD term, which suppresses a precession of magnetization and the non-adiabatic transition. As a result, the fidelity increases compared to the conventional QA.}
\label{fig11111}
\end{figure*}
Counter diabatic (CD) driving is one of the methods to address this issue \cite{Demirplak_2003, Berry_2009, Chen_2010, Campo_2013, Shuoming_2016, Guery-Odelin_2019}. 
In QA with CD driving, the ground state 
of the original time-dependent Hamiltonian
becomes the solution of the Schr\"{o}dinger dynamics.
The adiabatic dynamics can be attained using CD term even for a shorter computational time than conventional QA.

However, there are two difficulties in its construction for QA:  
(1) 
The CD term generally is non-local Hamiltonian,
and it is difficult to construct such non-local Hamiltonians in the actual experiments \cite{Campo_2012, Campo_2013}. 
(2) To construct the CD term, we need to know
the exact eigenstates of the quantum many-body Hamiltonian at all times.
The latter is more intractable
because we should not know the eigenstates of the Hamiltonian when we solve practical problems with QA.
Various methods have been devised to construct CD terms using the variational method and mean field (MF) approximation to tackle this problem
\cite{Sels_2017, Hatomura_2017, Claeys_2019, Prielinger_2021, Mbeng_2022, Kadowaki_2023}.

A method for constructing the CD term based on MF approximation has been proposed by Hatomura \cite{Hatomura_2017}. 
The advantages of MF approximation are that it simplifies the problem into an one-body problem and allows the CD term to be represented by local operators. 
In Ref. \cite{Hatomura_2017}, they dealed with a uniformly interacting fully connected ferromagnetic Ising model.
In the case of interaction with uniform ferromagnetic, 
the self-consistent equation in MF theory can be represented as the closed form of magnetization in the system. 
However, more interesting and practical problems arise when the interactions are antiferromagnetic or non-uniform, i.e., frustration and disorder play a crucial role. 
It is unclear whether the MF theory can be applied to such a general Hamiltonian.

The self-consistent equations show a significant difference between the case of uniform ferromagnetic interactions and the others. 
In systems with non-uniform interactions such as a spin glass model \cite{Edwards_1975, Sherrington_1975}, it is necessary to distinguish the magnetization of each qubit. In this case, the self-consistent equation becomes $N$ nonlinear simultaneous equations, where $N$ is the number of qubits. 
Furthermore, constructing the time-dependent MF approximated CD (MFCD) term requires solving $N$ nonlinear simultaneous equations at each time step in QA. 
There is no known method to
efficiently
solve these equations by using a classical computer.
Therefore, constructing MFCD can no longer be considered efficient for a general setup in the previous approach.

In this paper, we propose a general
and practical
method to construct the MFCD driving.
To address the difficulty of solving self-consistent equations, we construct the MFCD term by solving the dynamics of the magnetization \cite{Hatomura_2018} for each qubit by
using a classical computer (see Fig. 1).
Since the dynamics of magnetization can be described by classical equations of motion, it is possible to solve them using a classical computer in polynomial time with respect to the number of qubits $N$.
This allows us to construct MFCD term without solving self-consistent equations at every time step.
We apply this method to the spin glass model with transverse magnetic fields. We numerically
show that a fidelity of the ground state increases significantly compared to the conventional QA without the MFCD driving. 
Furthermore, we experimentally demonstrate 
our proposal by using the D-wave annealing machine,
and obtain experimental results that support these numerical calculations.

The remaining of our paper is as follows: In sec. II, we provides a review of QA and CD driving. In sec. III, we explain our methodology. In sec. IV, we describe numerical results. In sec. V, we explain the implementation and experimental results on D-Wave annealer. Sec. VI is devoted to the conclusion.

\section{Quantum annealing and CD driving}
Let us review QA and CD driving in this section.
\subsection{Quantum annealing}
In conventional QA, we consider the problem Hamiltonian where the ground state is a desired state
and we use the transverse magnetic field as the driver Hamiltonian to induce a quantum fluctuation. 
The QA Hamiltonian is given as follows \cite{Farhi_2000, Farhi_2001}. 
\begin{align}
    H_{0}(t)=f\left(\frac{t}{T}\right)H_{P}(t)+\left[1-f\left(\frac{t}{T}\right)\right] H_{D}(t),
\end{align}
where $H_{P}$, $H_{D}$, and $T$ denote the problem Hamiltonian, the driver Hamiltonian, and  the annealing time, respectively.
Also,
$f(t/T)$ is a continuous function with boundary condition such that $f(0)=0$ and $f(1)=1$.
Suppose that the initial state is a ground state of $H_{D}$ at $t=0$.
If
the adiabatic condition is satisfied, 
the system remains in the instantaneous ground state of $H_{0}(t)$ at each time,
and so we can obtain the ground state of $H_{P}$ at $t=T$.
Throughout our paper, the unit of the Hamiltonian is GHz, and that of the time is ns.
\subsection{Exact and MF approximated CD driving}
If the time evolution is adiabatic, a state driven by Hamiltonian (1) is given as
\begin{align}
\ket{\Psi_{n}(t)}=e^{i\gamma_{n}(t)}\ket{n(t)}.
\end{align}
Here, $\ket{n(t)}$ denotes instantaneous eigenstate of the Hamiltonian (1), and
$\gamma_{n}(t)$ denotes the phase factor in adiabatic dynamics, which is given as following,
\begin{align}
\gamma_{n}(t)=-\frac{1}{\hbar}\int_{0}^{t}dt'E_{n}(t')+i\int_{0}^{t}dt'\braket{n(t')|\partial_{t'}n(t')},
\end{align}
where $E_{n}$ denotes the $n$-th eigenvalue of $H_{0}(t)$.
We introduce the control field $H_{\rm cd}$ such that the eigenstate (2) becomes the solution of Schr\"{o}dinger equation. 
\begin{align}
i\frac{\partial}{\partial t}\ket{\Psi_{n}(t)}= [H_{0}(t)+H_{\rm cd}(t)]\ket{\Psi_{n}(t)},
\end{align}
where $H_{\rm cd}(t)$ is given by
\begin{align}
\label{eq5}
H_{\rm cd}(t)=i\sum_{n\neq m}\frac{\braket{\Psi_{n}(t)|\partial_{t}H_{0}(t)|\Psi_{m}(t)}}{E_{m}(t)-E_{n}(t)}\ket{\Psi_{n}(t)}\bra{\Psi_{m}(t)}.
\end{align}
This operator (\ref{eq5}) involves a non-local interaction which is difficult to experimentally implement. 
Therefore,
it is desirable
to represent $H_{\rm cd}(t)$ 
by using
local operators.
For special cases, it is known that
the MF approximation enables us to express $H_{\rm cd}(t)$ 
by using the local operators.
In Ref. \cite{Hatomura_2017}, they constructed the MFCD term for the uniform ferromagnetic Hamiltonian and its MF approximation 
is represented
as following
\begin{align}
\label{eq6}
H_{0}(t)=-f(t)\frac{J}{2N}\sum_{i,j}\hat{\sigma}^{z}_{i}\hat{\sigma}^{z}_{j}
-(1-f(t))\Gamma\sum_{i}\hat{\sigma}^{x}_{i}-h\sum_{i}\hat{\sigma}^{z}_{i},
\end{align}
\begin{align}
\label{eq7}
H^{\rm MF}_{0}(t)&=-f(t)\frac{JN}{2}(m^{z}(t))^2-f(t)(Jm^z(t)+h)\sum_{i}\hat{\sigma}^{z}_{i}\nonumber\\
&-(1-f(t))\Gamma\sum_{i}\hat{\sigma}^{x}_{i}.
\end{align}
where $\hat{\sigma}_{i}^{\alpha}$ ($\alpha=x$, $y$, $z$) are Pauli matrices at $i$th site, $J>0$, $\Gamma$, $h$ are the strength of interaction, transverse field, and longitudinal field, respectively.
The MFCD term can be constructed by using Eq. (\ref{eq5}) and (\ref{eq7})
as following,
\begin{align}
H^{\rm MF}_{\rm cd}(t)=B^{\rm eff}_{y}(t)\sum_{i}\hat{\sigma}_{i}^{y}, 
\end{align}
\begin{align}
B^{\rm eff}_{y}(t)=\frac{1}{2}\frac{-(Jm^{z}(t)+h)\dot{f}(t)\Gamma-J\dot{m}^z(t)(1-f(t))\Gamma}{(Jm^{z}(t)+h)^2+(1-f(t))^2\Gamma^2},
\end{align}
where $m^{z}(t)$ denotes the magnetization.  
In the MF theory, $m^{z}(t)$ is determined by solving the self-consistent equation:     
\begin{align}
\label{eq10}
m^{z}(t)=\braket{\Psi^{\rm MF}(t)|\hat{\sigma}^{z}|\Psi^{\rm MF}(t)},
\end{align}
where $\ket{\Psi^{\rm MF}(t)}$ is eigenstate of $H^{\rm MF}_{0}(t)$.
Solving the self-consistent equation (\ref{eq10}) is equivalent to the problem of searching for the zeros of an single-variable function.
In the case of the non-uniform interaction, 
the number of the
self-consistent equations 
is equal to
the number of qubits $N$.
This task is equivalent to the solving the non-linear simultaneous equation, which is not tractable by using a classical computer.
\section{Method}
We consider the following total Hamiltonian: 
\begin{align}
\label{eq11}
     H_{0}(t)=f(t)H_{P}+\left(1-f(t)\right)H_{D} 
     +g(t)H_{L}.
\end{align}
Here, we focus on the transverse Ising model as an example because of its simplicity. However, the following construction is not restricted to the transverse Ising model.
The problem Hamiltonian $H_{P}$ is given as
\begin{align}
H_{P}=-\sum_{i, j}J_{ij}{\hat \sigma}^{z}_{i}{\hat \sigma}^{z}_{j}-\Gamma\sum_{i}\hat{\sigma}^{x}_{i}, 
\end{align}
where $J_{ij}$ denotes the strength of interaction between $i$th and $j$th qubits.
We consider the driver Hamiltonian $H_{D}$ as the transverse magnetic field,
\begin{align}
H_{D}=-\Gamma_{D}\sum_{i}\hat{\sigma}^{x}_{i},
\end{align}
where $\Gamma_{D}$ denotes the strength of the transverse magnetic field of the driver Hamiltonian.
$H_{L}$ denotes the longitudinal magnetic field as following,
\begin{align}
H_{L}=-\sum_{i}h_{i}{\hat\sigma}^{z}_{i},
\end{align}
Here, $h_{i}$ is a uniform random longitudinal magnetic field for $i$th site, with $h_{i} \in [0, 1]$.
We should note that the classical motion of magnetization is often trapped in the local minima of the energy landscape, which prevents us from obtaining the true ground state \cite{Hatomura_2018}.
To avoid this problem, we adopt the inhomogeneous magnetic field.
It is worth mentioning that
we can control both transverse and longitudinal magnetic fields for each qubit in our framework.
However, to demonstrate our method with a D-wave QA device (sec. V), we 
assume that the longitudinal field is non-uniform, and the transverse field is uniform.

We choose the scheduling functions $f(t)$ and $g(t)$ such that MFCD term is zero at initial and final time.
We adopt the $f(t)$ and $g(t)$ as $f(t)=\frac{1}{2}[1-{\rm cos}(\pi t/T)]$, $g(t)=\frac{1}{2}\ {\rm sin}^2(\pi t/T)+\delta$, respectively.
Here, we introduce a parameter, $\delta$, which gives an infinitesimal field to resolve degeneracy of the energy spectrum at $t=T$. We set as $\delta=10^{-3}$.
The MF Hamiltonian of Eq. (\ref{eq11}) is given by
\begin{align}
\label{eq15}
    H_{0,i}^{\rm MF}(t)&=\left[\bigotimes_{j\neq i}{}_{j}\bra{\Psi^{\rm MF}(t)}\right]H_{0}(t)\left[\bigotimes_{j\neq i}\ket{\Psi^{\rm MF}(t)}_{j}\right]\nonumber\\
    &=-\left[f(t)\sum_{j\neq i}J_{ij}m^{z}_{j}(t)+g(t)h_{i}\right]{\hat \sigma}^{z}_{i}\nonumber\\
    &-\left[(1-f(t))\Gamma_{D}+\Gamma\right]{\hat \sigma}^{x}_{i},
\end{align}
where $m^{z}_{i}(t)$ is magnetization at $i$th site as 
\begin{align}
\label{eq16}
    m^{z}_{i}(t)={}_{i}\braket{\Psi^{\rm MF}(t)|\hat{\sigma}^{z}_{i}|\Psi^{\rm MF}(t)}_{i}.
\end{align}
In MF theory, the effective magnetic field $\bm{B}^{\rm eff}_{i}(t)$ is given as
\begin{align}
\label{eq17}
B^{\rm eff}_{z,i}(t)=f(t)\sum_{j\neq i}J_{ij}m^{z}_{j}(t)+g(t)h_{i},
\end{align}
\begin{align}
\label{eq18}
B^{\rm eff}_{x,i}(t)=(1-f(t))\Gamma_{D}+\Gamma,
\end{align}
and 
\begin{align}
\label{eq19}
B^{\rm eff}_{y,i}(t)=\frac{1}{2}\frac{\cfrac{ d B^{\rm eff}_{z,i}(t)}{dt} B^{\rm eff}_{x,i}(t)-\cfrac{ d B^{\rm eff}_{x,i}(t)}{dt} B^{\rm eff}_{z,i}(t)}{B^{\rm eff\ 2}_{z,i}(t)+B^{\rm eff\ 2}_{x,i}(t)}.
\end{align}
Here, Eq. (\ref{eq19}) denotes the magnetic field along $y$-axis which yields the MFCD drive.
We obtain the MFCD term as follows:
\begin{align}
\label{eq20}
H^{\rm MF}_{\rm cd}(t)=-\sum_{i}B^{\rm eff}_{y,i}(t)\hat{\sigma}^{y}_{i}
\end{align}
In Eq. (\ref{eq16}), $\ket{\Psi^{\rm MF}(t)}_{i}$ depends on the $m^{z}_{j}(t)$ ($j$ denotes all sites except for the $i$th site).
To construct the MFCD term (\ref{eq20}), Eq. (\ref{eq16}) has to be solved.
However, this simultaneous equation is non-linear, which is generally hard to solve.

To address this problem, we consider 
obtaining the time dependence of magnetization at each site by solving the
magnetization dynamics \cite{Hatomura_2018} using a classical computer instead of solving self-consistent equation (\ref{eq16}). 
In the next section, we numerically confirm that the solution of self-consistent equation is equivalent to the magnetization dynamics by using MFCD term.
It is well known that the magnetization dynamics is given by the Bloch equation, i.e., 
\begin{align}
\label{eq21}
\frac{d\bm{m}_{i}(t)}{dt}=2\bm{m}_{i}(t)\times \bm{B}^{\rm eff}_{i}(t).
\end{align}
Since $d B^{\rm eff}_{z,i}(t)/dt$ explicitly depends on $d m^{z}_{i}(t)/dt$, we can transform $d B^{\rm eff}_{z,i}(t)/dt$ into the following, 
\begin{align}
\label{eq22}
\frac{ d B^{\rm eff}_{z,i}(t)}{dt}&=\frac{d f(t)}{dt}\sum_{j\neq i}J_{ij}m^{z}_{j}(t)-f(t)\sum_{j\neq i}J_{ij}\frac{d m^{z}_{j}(t)}{dt}\nonumber\\
& +\frac{d g(t)}{dt}h_{i}\nonumber\\
&=\frac{d f(t)}{dt}\sum_{j\neq i}J_{ij}m^{z}_{j}(t)+\frac{d g(t)}{dt}h_{i}\nonumber\\
&+f(t)\sum_{j\neq i}J_{ij} m^{x}_{j}(t)\frac{B^{\rm eff}_{x,j}(t)}{B^{\rm eff\ 2}_{z,j}(t)+B^{\rm eff\ 2}_{x,j}(t)}\frac{d B^{\rm eff}_{z,j}(t)}{dt}\nonumber\\
&-f(t)\sum_{j\neq i}J_{ij} m^{x}_{j}(t)\frac{B^{\rm eff}_{z,j}(t)}{B^{\rm eff\ 2}_{z,j}(t)+B^{\rm eff\ 2}_{x,j}(t)}\frac{d B^{\rm eff}_{x,j}(t)}{dt}\nonumber\\
&+2f(t)\sum_{j\neq i}J_{ij}m^{y}_{j}(t)B^{\rm eff}_{x,i}(t).
\end{align}
Since Eq. (\ref{eq22}) is a set of
linear simultaneous equations regarding $d B^{\rm eff}_{z,i}(t)/dt$, we can easily obtain the solution numerically.
If we set the initial condition as
 $m^{x}_{i}(0)=1$, $m^{y}_{i}(0)=m^{z}_{i}(0)=0$, which corresponds to the initial state of classical transverse field, we solve classical equation of motion and then we obtain $m^{x}_{i}(t), m^{y}_{i}(t)$ and $m^{z}_{i}(t)$.
Therefore, we can obtain the entire time dependence of $H^{\rm MF}_{\rm cd}(t)$.
 
\section{Numerical result}
In this section, first, 
we numerically show the equivalence between the magnetization dynamics with MFCD term and the self-consistent equation of MF theory.
Subsequently, we show numerical results of QA with MFCD term.
We consider the time evolution of the ground state of $H_{D}$, $\ket{\Psi_{0}(0)}$ by following Schr\"{o}dinger equation:
\begin{align}
i\frac{\partial}{\partial t}\ket{\Psi_{0}(t)}= [H_{0}(t)+H^{\rm MF}_{\rm cd}(t)]\ket{\Psi_{0}(t)}.
\end{align}
The fidelity is defined as $|\braket{\Psi_{0}(t)|\phi(t)}|^2$, where $\ket{\phi(t)}$ denotes the ground state of Eq. (\ref{eq11}) with exact diagonalization at each time.
\begin{figure}[t]
\includegraphics[scale=0.72, bb=240 0 100 400]{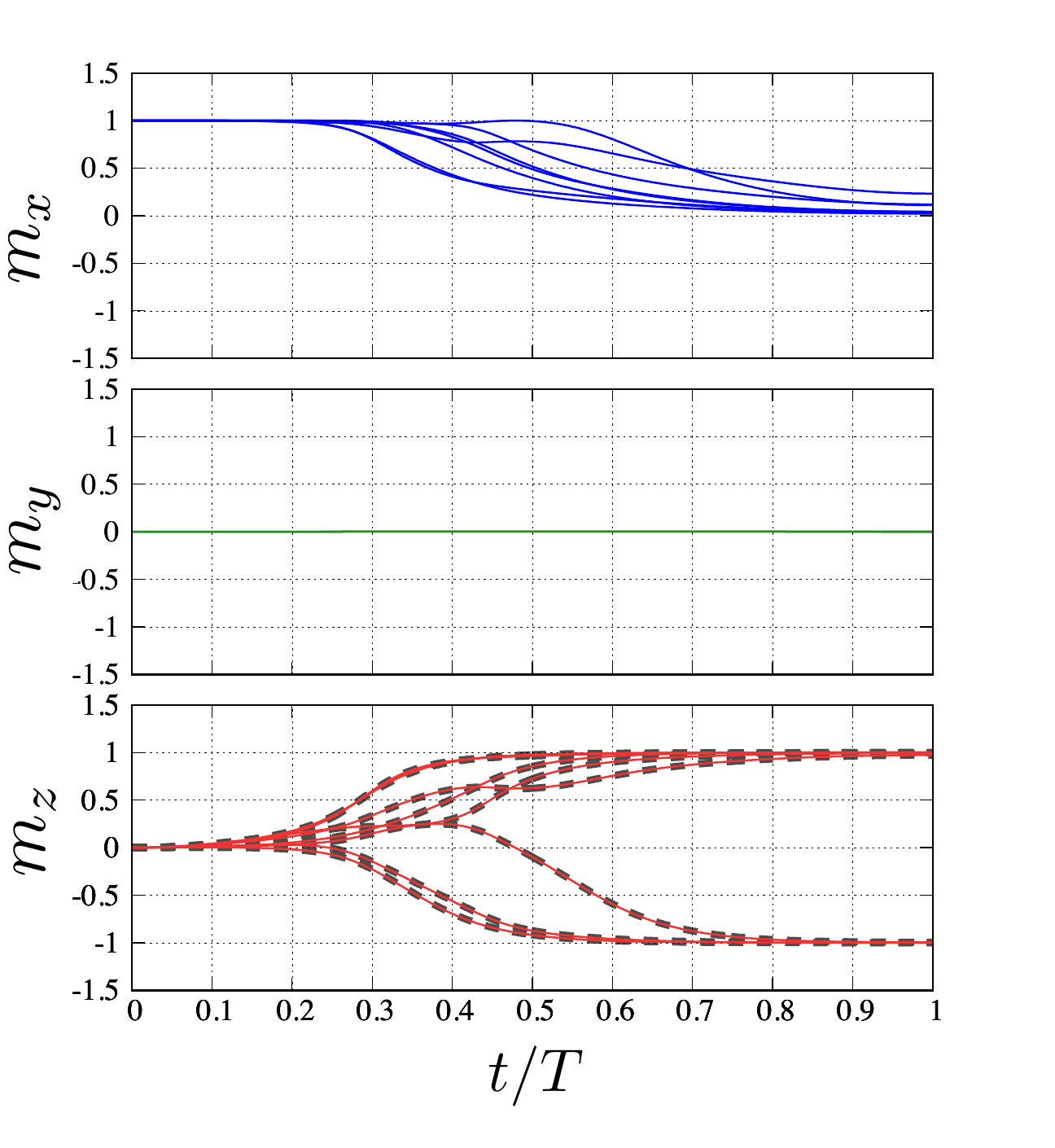}
\caption{The dynamics of magnetization in the fully connected spin glass model with random longitudinal and transverse field.
The solid line shows the magnetization at each qubit by solving Eq. (\ref{eq21}). The dotted line in the plot of $m_{z}$ shows the solution of self-consistent equation (\ref{eq16}). We set $T=1$,  $\Gamma=0.1$, $\Gamma_{D}=1$, and $N=8$.
}
\label{fig1}
\end{figure}
The coupling strength
$J_{ij}$ is
generated from the following distribution,
\begin{align}
\label{eq24}
P(J_{ij})\sim \frac{1}{\sqrt{2\pi \sigma^{2}}}{\rm exp}\left(-\frac{J^2_{ij}}{2\sigma^2}\right),
\end{align}
where, $\sigma$ denotes the variance, and we set $\sigma=1$.

\subsection{Dynamics of magnetization}
In this subsection, we consider a fully connected spin glass model \cite{Sherrington_1975} with the transverse field as the QA Hamiltonian.
Fig. \ref{fig1} shows the dynamics of magnetization (\ref{eq21}) in a certain 
 set of random interaction \{$J_{ij}$\} and random fields \{$h_{i}$\}. 
Due to the random interaction and the longitudinal magnetic field, the magnetization at each site undergoes non-uniform rotation.
In the case without MFCD term ($B^{\rm eff}_{y,i}(t)=0$), 
the motion of magnetization is different from the
solution of 
the
self-consistent equations (\ref{eq16}) because of a precession around the effective magnetic field.
On the other hand, in our method, the magnetization tracks
that of
the instantaneous stationary state due to the
MFCD terms (\ref{eq19}).
Actually, in Fig. \ref{fig1}
 $m^{y}_{i}(t)=0$
is satisfied, which means that
the motion of magnetization is constrained in $x$--$z$ plane. 
Moreover, in Fig. \ref{fig1}, 
we show that the solution of Eq. (\ref{eq16}) at each time is consistent with the dynamics of magnetization with MFCD term (\ref{eq21}).
Since we cannot efficiently solve the equation (\ref{eq16}) for the large number of spins by using a classical computer,
it is noteworthy that we can solve the dynamics with MFCD term,
which corresponds to  the self-consistent equation (\ref{eq16}).
It should be noted that we have checked 30 samples for $J_{ij}$ and 100 samples for $h_{i}$, which exhibits 
the correspondence between the self-consistent equation (\ref{eq16}) and the magnetization dynamics (\ref{eq21}) in almost all samples. However, for a few samples, the magnetization dynamics does not track the instantaneous stationary state due to criticality in classical spin systems \cite{Hatomura_2018}.
\subsection{Performance of QA}
\begin{figure*}[t!]
\includegraphics[scale=0.65, bb=750 0 0 300]{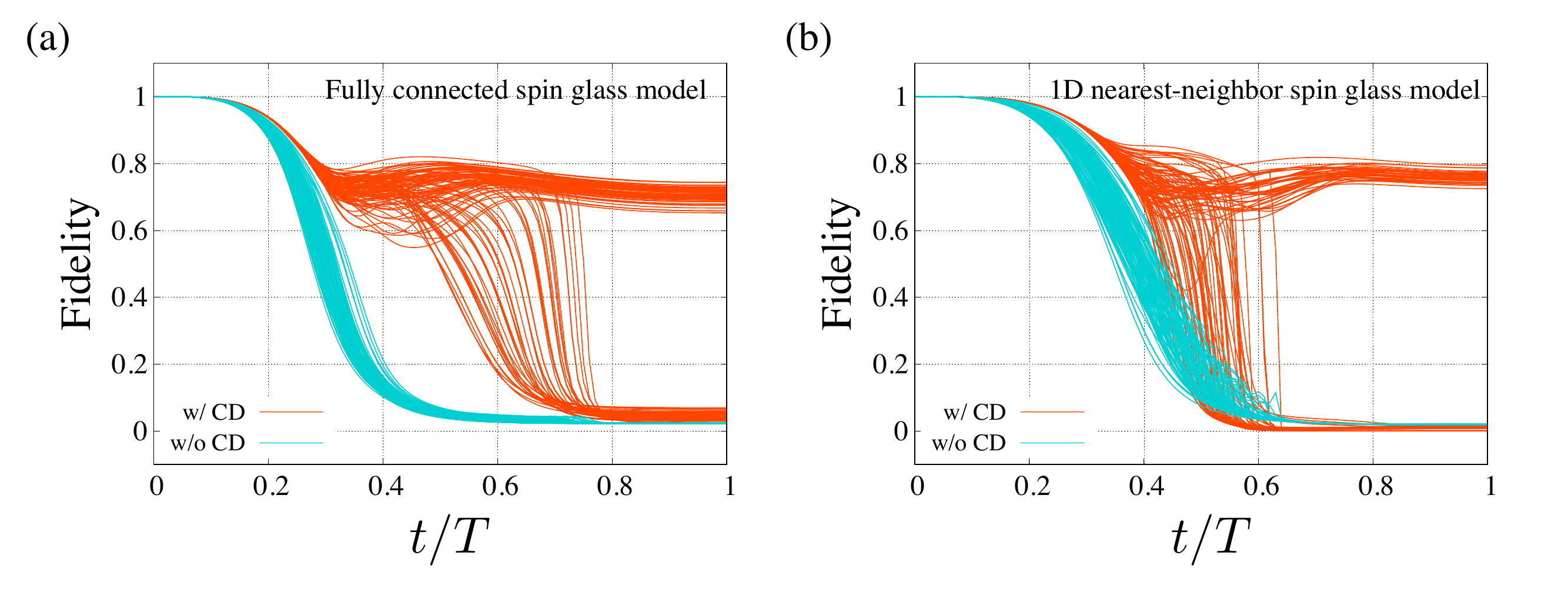}
\caption{The fidelity in the spin glass model with transverse field.
(a) and (b) show the case with fully connected and nearest neighbor interaction, respectively. 
In fixed $J_{ij}$, we plot for all cases with 100 samples of longitudinal field each, with and without MFCD term. 
The red and blue lines show the case with and without MFCD term, respectively. We set $T=1$,  $\Gamma=0.1$, $\Gamma_{D}=1$, and $N=8$.}
\label{fig4}
\end{figure*}

In this subsection, we show the numerical simulation of QA.
We consider the fully connected spin glass model
\cite{Sherrington_1975}
and
one dimensional model with nearest-neighbour interactions \cite{Edwards_1975}.
It is generally expected that
as the dimension increases,
fluctuations are suppressed and the MF approximation becomes more accurate \cite{Ginzburg_1961, Harris_1976}.
From this perspective, 
it may be expected that
the MF approximation will not be valid for the 1D spin chain in the limit of a large number of spins.
However, for the finite size system, the validity of MF  approximation is non-trivial.

Fig. \ref{fig4} (a) and (b) show a typical result of fidelity in the case of fully connected and 1D spin chain, respectively.
As shown in Fig. \ref{fig4} (a), in the fully connected spin glass, the decrease of fidelity is suppressed
due to the MFCD terms
in many samples.
Surprisingly, as shown in Fig. \ref{fig4} (b), the decrease of fidelity is also suppressed
due to the MFCD terms for the case of 1D spin chain.
Therefore, this result indicates that our MFCD term can be effective or successful even for the finite range interaction and transverse field.
It should be noted that we have checked 30 samples of $J_{ij}$, which exhibits almost same behavior. However, for a few samples,
the fidelity in all realization of the longitudinal field 
is around $0.7$
or nearly $0$. 
The range and number of samples for exploring the longitudinal fields will be addressed in future work.

\begin{figure}[t!]
\includegraphics[scale=0.38, bb=520 0 100 790]{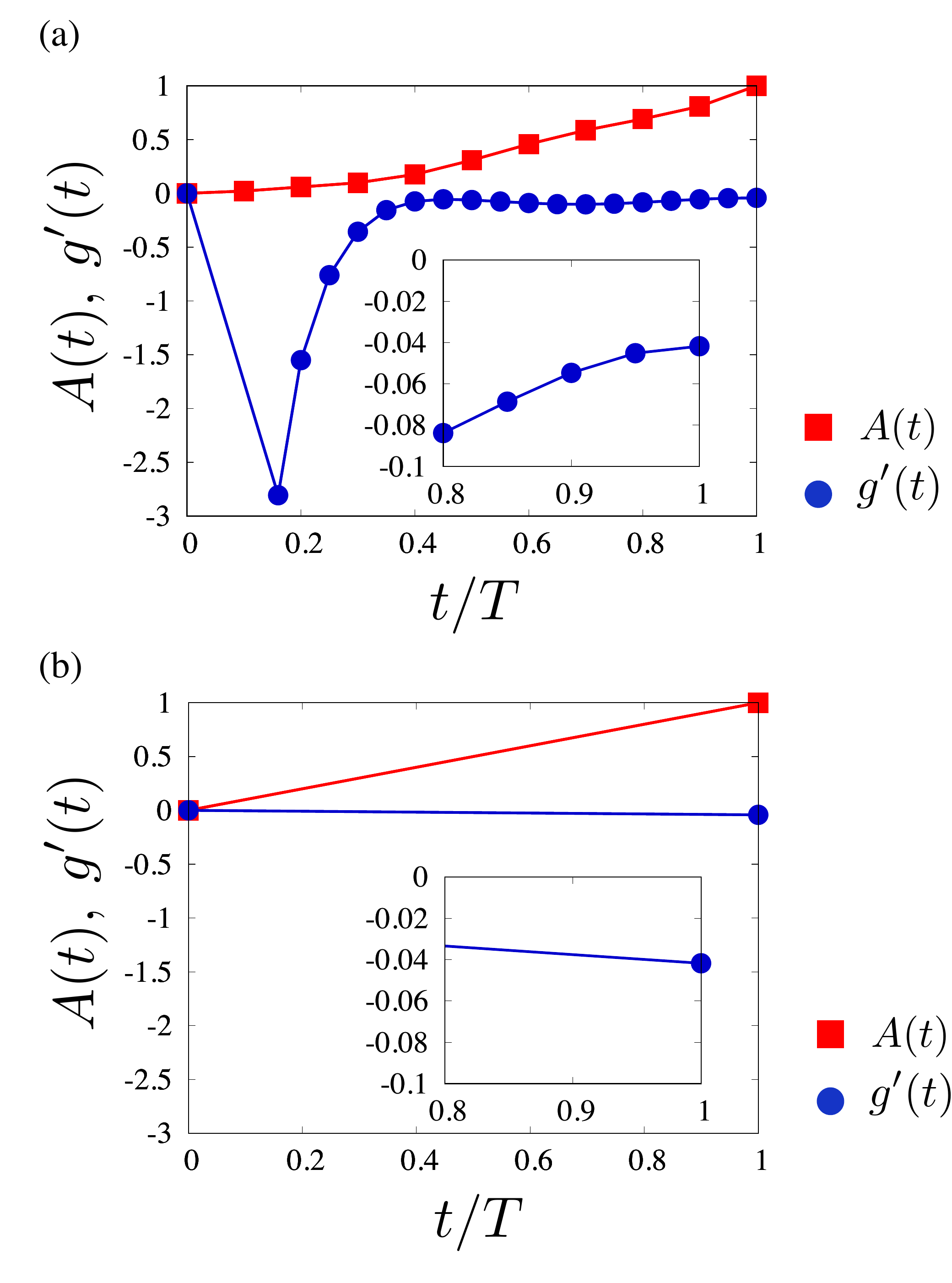}
\caption{Scheduling function of QA $A(t)$ and longitudinal field $g'(t)$ in the rotational frame used in our experiment. 
(a) and (b) show the case with MFCD and linear schedule, respectively.
Insets show the scheduling function of longitudinal field $g'(t)$ around $t=T$.}
\label{fig5}
\end{figure}
\begin{figure}[h!]
\includegraphics[scale=0.38, bb=520 10 100 480]{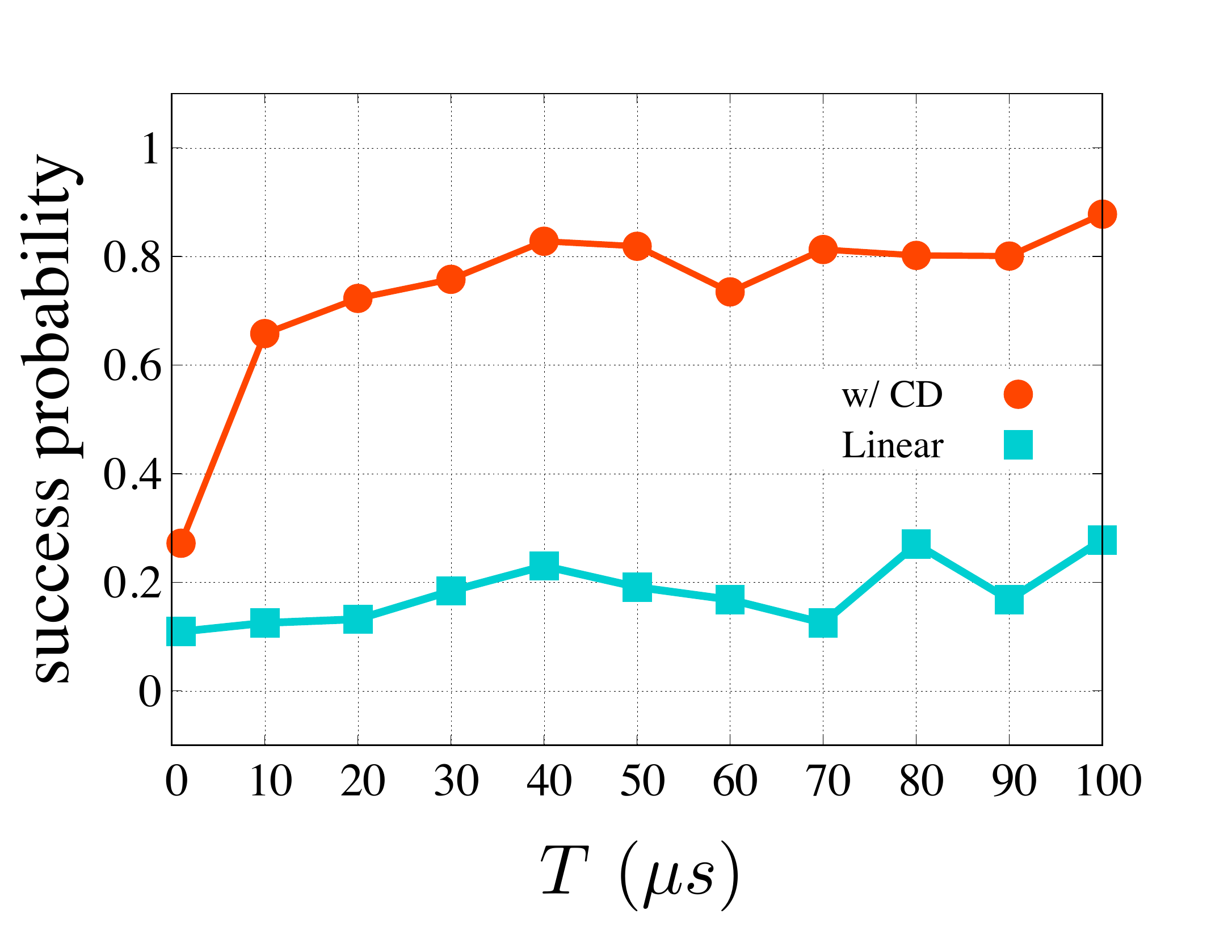}
\caption{The success probability in the fully connected anti-ferromagnetic model using D-Wave Advantage System 4.1 as a function of annealing time $T$. 
The red and blue line show the case with MFCD term and liner schedule, respectively.
}
\label{fig6}
\end{figure}
\section{EXPERIMENT USING D-WAVE QA MACHINE}
Finally, we experimentally perform QA on
a
D-wave QA machine as a demonstration of our method.
We use a D-Wave Advantage system 4.1.
We 
change the basis
to erase $\hat{\sigma}_{y}$ term
in the Hamiltonian
because we cannot use
$\hat{\sigma}_{y}$ on
the
D-wave 
 annealer.
 This technique has been used in previous works
 \cite{Sels_2017, Prielinger_2021,Kadowaki_2023}. 
The time-dependent rotational operator around $z$-axis
is
given by
\begin{align}
U(t)={\rm exp}\left(\frac{i}{2} \sum_{i} \phi_{i}(t)\hat{\sigma}_{i}^{z}  \right). 
\end{align}
In the rotating frame, $H_{0}(t)+H^{\rm MF}_{\rm cd}(t)$ is given by
\begin{align}
\label{eq26}
H_{rot}(t)&=f(t)H_{P}-\frac{1}{2}\sum_{i}\frac{d\phi_{i}(t)}{dt}\hat{\sigma}_{i}^{z}\nonumber\\
 &+g\left(t\right)H_{L} -\sum_{i}\sqrt{(1-f(t))^2+B^{\rm eff}_{y,i}(t)^2}\hat{\sigma}^{x}_{i}, 
\end{align}
where $\phi_{i}(t)$ is $\phi_{i}(t)={\rm Tan}^{-1}(B^{\rm eff}_{y,i}(t)/(1-f(t)))$.
As shown in
Eq. (\ref{eq26}), 
if the interaction or the longitudinal field have random values, 
independent control of the longitudinal and transverse fields is necessary for each qubit.
However, in the current D-wave QA machine,
we cannot change individual schedule of longitudinal fields but can change
the sign and coefficient
of longitudinal fields
for each qubit. 
Therefore, for simplicity,
we consider fully connected antifferomagnetic model without transverse magnetic field ($\Gamma=0$)  given by $J_{ij}=-1<0$ (for all $i$, $j$) \cite{chandra_2010}
where the problem Hamiltonian is represented as
$H_{P}=\sum_{i,j}\hat{\sigma}^{z}_{i}\hat{\sigma}^{z}_{j}$.
This system has degenerate ground states in which half of the spins are up and the other half are down when the number of spins is even.
The number of the ground states is given by $_NC_{N/2}$.
If we add the random longitudinal field, $h_{i} \neq$ 0 at $t=T$, there is a unique ground state which slightly apart from excited states in this system.
Therefore, it can be expected that this model has well captured a frustrated nature, and solving such a problem Hamiltonian is difficult with conventional QA.

As the random field, we
consider the following case
\begin{align}
\label{eq27}
h_{i} = 
\Bigg\{
\begin{array}{ll}
+h & (i\ {\rm for}\ {even})\\
-h & (i\ {\rm for}\ {odd})
\end{array}
,
\end{align}
where we set $h=1.1$. In this longitudinal field (\ref{eq27}), the classical N\'eel state, $\ket{\downarrow\ \uparrow\ \downarrow\ \uparrow\ \downarrow\ \uparrow\ \downarrow\ \uparrow\ }$ is realized as the ground state.
The Hamiltonian of the
D-wave QA machine
is given as follow:
\begin{align}
H_{0}&(t)= \Biggl[A(t)\Biggl(\frac{1}{3}H_{P}+4g'(t)\sum_{i}\frac{h_{i}}{h}\hat{\sigma}_{i}^{z}\Biggr)-\frac{1}{3}B(t)\sum_{i}\hat{\sigma}_{i}^{x} \Biggr].
\end{align}
Since D-wave QA machine can not control $A(t)$ and $B(t)$ independently, we assume $A(t)$ has a form as $A(t)=1-B(t)$.
In order to obtain this form, we divide Eq. (26) by $f(t)+\sqrt{(1-f(t))^2+B^{\rm eff}_{y}(t)^2}$, 
then we have
\begin{align}
&A(t)=\frac{f(t)}{f(t)+\sqrt{(1-f(t))^2+B^{\rm eff}_{y}(t)^2}},\\
&g'(t)=-\frac{g(t)+\frac{1}{2}\frac{d\phi(t)}{dt}}{12A(t)},\\
&B(t)=\frac{\sqrt{(1-f(t))^2+B^{\rm eff}_{y}(t)^2}}{f(t)+\sqrt{(1-f(t))^2+B^{\rm eff}_{y}(t)^2}},
\end{align}
where we replace $B^{\rm eff}_{y,i}$ and $\phi_{i}(t)$ with  $B^{\rm eff}_{y}$ and $\phi(t)$, respectively, since these factor is now uniform.
The scheduling function on
the D-wave QA machine is represented by several discretized intervals. 
Fig. \ref{fig5} (a) shows the partition points of the annealing and longitudinal field scheduling. 
Since we cannot start from a finite value of the longitudinal field, in the first interval, 
the longitudinal field linearly increase from 0 to the maximum range of that.
In D-Wave Advantage system 4.1, the range of the longitudinal field scheduling $g(t)'$ is $[-3.0, 3.0]$.
Here, Eq. (28) is multiplied by 1/3 to rescale $g'(t)$.
To compare the performance of QA without MFCD terms with that with MFCD terms,
we use a linear schedule shown in Fig. \ref{fig5} (b).
We implement
QA by varying the annealing time from $T=1\ \mu s$ to 100$\ \mu s$ and 
perform 1000 measurements
at each $T$. 
It should be noted that the fidelity in the transverse Ising model can not be directly measured because the quantum tomography can not be implemented by the D-wave QA machine. 
Therefore, we perform a measurement in computational basis for Ising model without transverse field, then we evaluate a performance of QA by using a success probability for obtaining the ground state of the Ising model.   
The success probability is defined as a ratio to obtain N\'eel state 
within 1000 trials.
Fig. \ref{fig6} clearly shows that the success probability with MFCD term 
is much higher than that
without MFCD term for all $T$.
This result is consistent with our result of numerical simulations that 
we
can prepare the target
ground state of 
the problem Hamiltonian 
with a high fidelity
by using QA with MFCD drive.
\\
\section{Conclusion}
In this paper, we proposed QA using MFCD term.  
Specifically, by solving classical motion of magnetization, we constructed the MFCD term.
We numerically
showed that the classical motion of magnetization with MFCD term is equivalent to solution of the self-consistent equation of the MF theory.

By using this MFCD term, we evaluated the performance of QA for ground state search 
in the spin glass model with transverse field. 
Performing the numerical simulation of QA in this model, 
we showed that the fidelity in QA with MFCD term 
is higher
in most samples than that without MFCD term. 
We obtain the enhancement of the fidelity
not only for fully connected models but also for 1D spin chains.
Moreover, we experimentally demonstrated QA with MFCD drive on the D-wave annealer, 
and
we showed that the success probability significantly increases by using our method.

In this paper, although
we used the spin glass model with transverse field as an emblematic case of a hard problem, 
our framework can be also
applied to the quantum Heisenberg model and other models even with such $XX$, $YY$, or more than two-body interaction.  
Therefore, our approach has the potential to become a general method
to improve
the efficiency of the quantum simulation using QA machines.
\begin{acknowledgments}
We would like to thank Tadashi Kadowaki for insightful discussions.
This paper was based on results obtained from a project, JPNP16007, commissioned by the New Energy and Industrial Technology Development Organization (NEDO), Japan.
This work was also supported by the Leading Initiative for Excellent Young Researchers, MEXT, Japan, and JST Presto
(Grant No. JPMJPR1919), Japan.
 This
work was supported by JST Moonshot R\&D (Grant
Number JPMJMS226C).
\end{acknowledgments}
\bibliography{apsbib}
\end{document}